\documentclass[lettersize,journal]{IEEEtran}
\usepackage{amsmath,amsfonts}

\usepackage{cite}
\usepackage{array}

\usepackage{tabularx}
\usepackage{booktabs}

\usepackage{bbding}
\usepackage[caption=false,font=normalsize,labelfont=sf,textfont=sf]{subfig}
\usepackage{multirow}
\usepackage{makecell}
\usepackage{textcomp}
\usepackage{stfloats}
\usepackage{url}
\usepackage{verbatim}
\usepackage{graphicx}
\usepackage[T1]{fontenc} 
\usepackage{amssymb} 
\usepackage{algorithmic}
\usepackage[ruled, vlined, linesnumbered]{algorithm2e}
\usepackage{threeparttable}
\usepackage[backref=False]{hyperref}
\def\BibTeX{{\rm B\kern-.05em{\sc i\kern-.025em b}\kern-.08em
    T\kern-.1667em\lower.7ex\hbox{E}\kern-.125emX}}
\usepackage{balance}
\newcommand{\Rmnum}[1]{\uppercase\expandafter{\romannumeral #1}}  
\usepackage{tikz,xcolor,hyperref}

\usepackage{amsmath}
\usepackage{subcaption}
\usepackage{graphicx}
\usepackage{amssymb}

\definecolor{lime}{HTML}{A6CE39}
\DeclareRobustCommand{\orcidicon}{
	\begin{tikzpicture}
		\draw[lime, fill=lime] (0,0)
		circle[radius=0.16]
		node[white]{{\fontfamily{qag}\selectfont \tiny \.{I}D}}; 
	\end{tikzpicture}
	\hspace{-2mm}
}
\foreach \x in {A, ..., Z}{%
	\expandafter\xdef\csname orcid\x\endcsname{\noexpand\href{https://orcid.org/\csname orcidauthor\x\endcsname}{\noexpand\orcidicon}}
}

\begin{document}
\title{Lightweight Low-SNR-Robust Semantic Communication System for Autonomous Driving}

\author{Ruixing~Ren\hspace{-1.5mm}\orcidA{}, Minjie~Wei, Junhui~Zhao\hspace{-1.5mm}\orcidC{}\IEEEmembership{Senior~Member,~IEEE}

\thanks{
(Corresponding author: Junhui Zhao.)
		

		
}
}

\maketitle

\begin{abstract}
  Image transmission for vehicle-to-vehicle collaborative perception in autonomous driving faces challenges including limited on-board terminal resources, time-varying wireless channel fading, and poor robustness under low signal-to-noise (SNR) ratio. Traditional separate source-channel coding schemes suffer from the cliff effect, while existing semantic communication models are limited by large parameter sizes and weak digital compatibility. This paper proposes a lightweight, low-SNR-robust deep joint source-channel coding (JSCC) semantic communication system. First, structured pruning is implemented based on batch normalization layer scaling factors and L1 regularization, which significantly reduces model complexity while ensuring image reconstruction quality. Second, a uniform quantization and M-QAM modulation scheme adapted to JSCC features is designed, and a training-deployment separation strategy is adopted to address the non-differentiable quantization problem, enabling compatibility with existing digital communication systems. Simulation results on the Cityscapes dataset show that the pruned model maintains comparable performance and robustness to the original one, even with over half of its parameters removed.
  Notably, the proposed scheme exhibits significant advantages over conventional communication methods under low SNR conditions.
\end{abstract}

\begin{IEEEkeywords}
Autonomous driving, semantic communication, collaborative perception, model pruning, image communication
\end{IEEEkeywords}

\section{Introduction}
With the rapid development of intelligent connected vehicles and internet of vehicles technologies, autonomous driving has gradually evolved from single-vehicle intelligence to a new stage of multi-vehicle collaborative perception and swarm intelligence decision-making \cite{RenITS,RenTVTIoV}. Vehicle-to-vehicle collaborative perception effectively breaks through line-of-sight constraints, occlusion blind spots, and adverse weather interference of single-vehicle sensors by sharing real-time environmental visual information among vehicles\cite{RenUAV,Shi}. It significantly improves environmental perception, decision reliability, and driving safety of autonomous driving systems, making it one of the key technologies enabling high-level autonomous driving. However, vehicular communication is constrained by time-varying wireless channel fading, low signal-to-noise ratio (SNR), limited bandwidth, and insufficient computation and storage resources of on-board terminals \cite{Yaoyu}. Traditional digital communication schemes based on separate source-channel coding can hardly meet the high-reliability, low-latency, and lightweight image transmission requirements of collaborative perception \cite{9852388}.

In traditional communication architectures, source coding and channel coding are designed independently based on the Shannon separation theorem. Source coding achieves high compression ratios by removing spatial redundancy. However, in low-SNR scenarios, channel noise easily causes bit error propagation, leading to decoding failure and a catastrophic drop in image reconstruction quality, known as the cliff effect \cite{2026arXiv260404413H}. Although channel coding improves anti-interference ability, it introduces extra computation overhead and transmission latency, which cannot adapt to resource constraints of vehicular edge computing platforms .
Besides, traditional schemes transmit raw pixel-level data without fully exploiting semantic priors and perception task requirements in autonomous driving scenarios, making it difficult to balance transmission efficiency and robustness \cite{10695151}.

Recently, as a novel paradigm breaking the performance limits of traditional communications, semantic communication has attracted extensive attention from both academia and industry. Instead of transmitting redundant bit-level data, semantic communication extracts and transmits task-critical semantic features via deep learning (DL) models \cite{9852388}. As a mainstream implementation of semantic communication, deep joint source-channel coding (JSCC) optimizes source and channel coding jointly via end-to-end neural networks \cite{8723589}. It effectively suppresses noise interference and avoids the cliff effect, demonstrating superior performance in image and video transmission tasks.

Recently, extensive studies have applied deep learning to semantic communication for image transmission. The work of \cite{9914635} proposed an end-to-end image communication system based on deep neural networks and built a field-programmable gate array-based verification platform, demonstrating its practical feasibility. Furthermore, for V2V communication scenarios, a Swin Transformer-based semantic communication system for image segmentation has been proposed. Experiments show that it outperforms conventional coding schemes in terms of mean intersection over union \cite{10118717}.  The work of \cite{10049005} employs a convolutional autoencoder to extract semantic features, with the base station providing decision support for connected autonomous vehicles upon receiving semantic information.
Furthermore, a general intelligent framework has been proposed to enable more efficient transmission and support various intelligent tasks in vehicular networks \cite{10013090}. This framework fuses relevant information from different users and achieves significantly higher rank-1 accuracy than traditional methods in cooperative identity retrieval for vehicular networks.

However, most existing semantic communication systems for image transmission rely on separate architectures or suffer from complex structures and large parameter sizes. According to \cite{8723589}, deep JSCC models show great potential in fading channels, especially under low SNR conditions. The work of \cite{9746335} noted that although most modern wireless image transmission systems were based on the Shannon separation theorem, its optimality held only when codeword length and complexity approach infinity. In scenarios with non-ergodic source and channel distributions and limited transmission bits, separate designs are suboptimal since they neglect the interaction between source coding and channel coding \cite{10015684}. Moreover, the growing demands of complex artificial intelligence tasks in intelligent transportation systems have exceeded the capacity of existing communication and computing resources. Thus, the allocation and scheduling of computational resources must be prioritized when deploying and executing models on vehicular computing platforms \cite{RenRIS}. Due to the complex architectures of some encoder-decoders, such systems may fail to meet the low-latency, high-reliability requirements of vehicular network applications.

Existing semantic communication systems usually feed real-valued outputs directly encoded by neural networks into the channel, which is incompatible with current digital communication systems \cite{10039447}. A major challenge in implementing digital modulation for semantic communication systems lies in the difficulty of optimizing neural networks via stochastic gradient descent. Some studies have explored modulation for semantic signals. For instance, a joint coding and modulation framework based on variational autoencoders has been proposed, which resolves the non-differentiability issue of digital modulation by learning the transition probability from source data to discrete constellation symbols \cite{bo2024joint}.

Although Deep JSCC performs excellently in general image transmission, its application to autonomous driving V2V cooperative perception still faces two key challenges:
\begin{enumerate}
	\item Existing JSCC codecs are mostly built on deep convolutional neural networks with large parameters and computational complexity, making them difficult to deploy directly on vehicle-mounted terminals with limited computing resources.
	\item Traditional semantic communication mostly adopts analog transmission schemes, which deliver excellent performance but are incompatible with existing digital communication systems.
	However, digital quantization modulation tends to introduce errors, leading to severe performance degradation under low SNR.
\end{enumerate}

To address the above issues, this paper focuses on the V2V collaborative perception scenario in autonomous driving and proposes a lightweight semantic communication system with low-SNR robustness. The main contributions of this paper are as follows:
\begin{itemize}
	\item A structured pruning method based on BN layer scaling factors and L1 regularization is proposed , which significantly reduces the parameter count and computational complexity of the JSCC model while ensuring image reconstruction quality, adapting to the resource constraints of vehicle-mounted terminals.
	\item A uniform quantization and M-QAM modulation transmission scheme adapted to JSCC encoded features is designed. A train-deploy separate strategy is adopted to solve the non-differentiable quantization problem, enabling compatibility with existing digital communication systems, while transmission robustness under low-SNR is significantly improved and the cliff effect is eliminated.
	\item Simulation experiments on the Cityscapes dataset verify the performance of the proposed system under various pruning rates, modulation orders and SNRs. Results show that the proposed scheme outperforms the conventional separate scheme in both lightweightness and robustness, meeting the practical transmission requirements for autonomous driving V2V collaborative perception.
\end{itemize}

The rest of this paper is organized as follows. Section \ref{2} establishes the V2V semantic communication model. Section \ref{3} elaborates on the lightweight pruning and low-SNR robust quantization-modulation scheme. Section \ref{4} verifies the proposed scheme performance via simulations. Section \ref{5} concludes the paper.

\begin{figure}[t]
	\centerline{\includegraphics[width=3.6in,keepaspectratio]{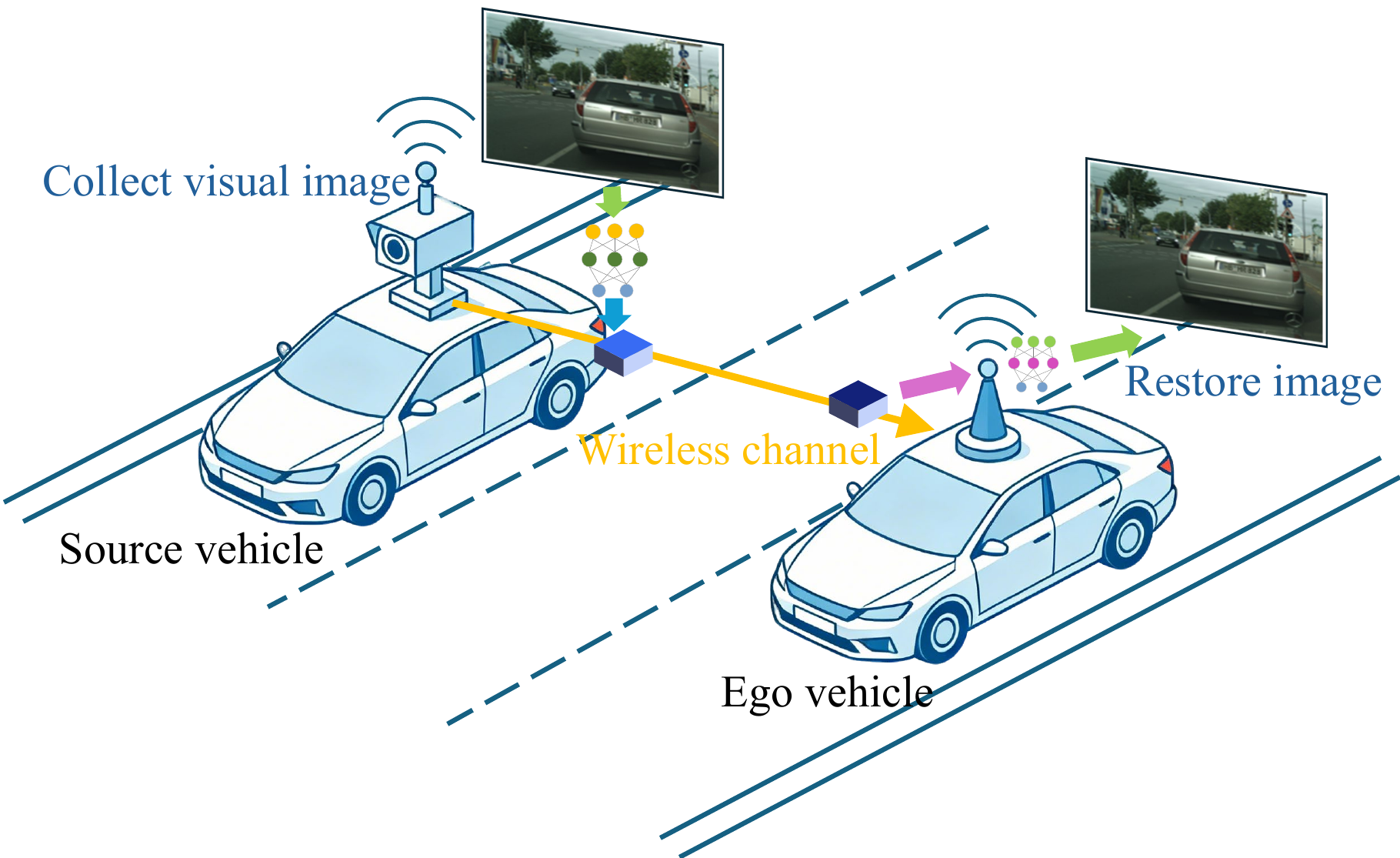}}
	\caption{Block Diagram of V2V Semantic Communication System for Autonomous Driving}
	\label{fig1}
\end{figure}
\section{System Model} \label{2}
This paper focuses on V2V collaborative perception tasks in autonomous driving scenarios. As shown in Fig. \ref{fig1}, visual images captured by the source vehicle are required to be transmitted to the nearby ego vehicle through wireless channels. The ego vehicle reconstructs images from received signals to support subsequent driving decisions. Given the limited computing and storage resources of on-board terminals and the time-varying fading characteristics of channels, the system must achieve low-complexity, robust transmission while maintaining high reconstruction quality.

Consider a single-antenna V2V link. Let the raw image captured by the source vehicle be $\mathbf{x} \in \mathbb{R}^{n}$, where $n = H \times W \times C$ denotes the source signal dimension, with $H$, $W$, and $C$ representing image height, width, and channel count, respectively. At the transmitter, a deep JSCC encoder first extracts the semantic features of the image, which can be expressed as:
\begin{equation}
	\mathbf{z}=\mathcal{E}_\theta (\mathbf{x}),
\end{equation}
where $\mathcal{E}_{\boldsymbol{\theta}}(\cdot)$ denotes the JSCC encoder with parameter set $\boldsymbol{\theta}$, and $\mathbf{z} \in \mathbb{R}^{k}$ is the extracted semantic feature vector. Here, $k$ denotes both the length of $\mathbf{z}$ and the number of symbols transmitted over the channel after quantization and modulation, also known as the number of channel uses. The bandwidth compression ratio is defined as $k/n$, which typically satisfies $k \ll n$ for efficient transmission. To ensure compatibility with existing digital communication systems and facilitate hardware implementation, the last layer of the encoder uses the Sigmoid activation function to constrain the output $\mathbf{z}$ to $(0,1)$.

Subsequently, $\mathbf{z}$ is quantized and modulated before being fed into the digital channel. Specifically, each element $z_i$ is uniformly quantized to $M$-level indices, which are then mapped to $M$-QAM constellation points $\mathbf{s}$.

The modulated symbols $\mathbf{s}$ are transmitted over a slow Rayleigh fading channel impaired by additive white Gaussian noise. The received signal can be expressed as:
\begin{equation}
	\mathbf{y} = \mathbf{h}\odot\mathbf{s} + \mathbf{n},
\end{equation}
where $\mathbf{h}$ denotes the complex channel gain coefficients, each following a Rayleigh distribution with zero mean and unit variance. $\mathbf{n} \sim \mathcal{CN}(0, \sigma^2 \mathbf{I})$ denotes complex AWGN, with variance $\sigma^2$ determined by the SNR. $\odot$ represents the Hadamard product. The transmit power constraint is enforced by normalizing the average power of $\mathbf{s}$, i.e.,
\begin{equation}
	\frac{1}{k}\mathbb{E}[|\mathbf{s}|_2^2] \le P.
\end{equation}

At the receiver, minimum distance demodulation is first performed on $\mathbf{y}$ to recover the quantization indices, yielding the quantized semantic feature $\hat{\mathbf{z}}$. Finally, the JSCC decoder $\mathcal{D}_{\boldsymbol{\phi}}(\cdot)$ with parameter set $\boldsymbol{\phi}$ reconstructs $\hat{\mathbf{z}} $ as
\begin{equation}
	\hat{\mathbf{x}}=\mathcal{D}_{\boldsymbol{\phi}}(\hat{\mathbf{z}}).
\end{equation}
The goal of the communication system is to make the reconstructed image $\hat{\mathbf{x}}$ as close as possible to the original image $\mathbf{x}$. Similarly, the last layer of the decoder also adopts the Sigmoid activation function, so that the output pixel values can reconstruct the original image by multiplying by 255 \cite{8723589}.

Peak signal-to-noise ratio (PSNR) and structural similarity (SSIM) are adopted as performance metrics to evaluate reconstruction quality in terms of pixel-level error and human-perceived structural similarity, respectively. PSNR is defined as:
\begin{equation}
	\mathrm{PSNR}(\mathbf{x},\hat{\mathbf{x}})=10\log_{10}\frac{\mathrm{MAX}^2}{\mathrm{MSE}},
\end{equation}
where $\text{MAX} = 255$ is the maximum pixel value of 8-bit images, and MSE denotes the mean squared error.
SSIM comprehensively assesses the similarity between two images in terms of luminance, contrast, and structure, and is defined as \cite{1284395}:
\begin{equation}
	\mathrm{SSIM}(\mathbf{x},\hat{\mathbf{x}})=l(\mathbf{x},\hat{\mathbf{x}})\cdot c(\mathbf{x},\hat{\mathbf{x}})\cdot s(\mathbf{x},\hat{\mathbf{x}}),
\end{equation}
where the luminance similarity $l(\mathbf{x},\hat{\mathbf{x}})$, contrast similarity $c(\mathbf{x},\hat{\mathbf{x}})$, and structural similarity $s(\mathbf{x},\hat{\mathbf{x}})$ are given by:
\begin{equation}
	l(\mathbf{x},\hat{\mathbf{x}})=\frac{2\mu_x\mu_{\hat{x}}+C_1}{\mu^2_x+\mu_{\hat{x}}^2+C_1},
\end{equation}
\begin{equation}
	c(\mathbf{x},\hat{\mathbf{x}})=\frac{2\sigma_x\sigma_{\hat{x}}+C_2}{\sigma^2_x+\sigma_{\hat{x}}^2+C_2},
\end{equation}
\begin{equation}
	s(\mathbf{x},\hat{\mathbf{x}})=\frac{\sigma_{x\hat{x}}+C_3}{\sigma_x\sigma_{\hat{x}}+C_3},
\end{equation}
In practice, $C_3=C_2/2$ is commonly adopted, allowing the three expressions to be simplified as:
\begin{equation}
	\mathrm{SSIM}(\mathbf{x},\hat{\mathbf{x}})=\frac{(2\mu_x\mu_{\hat{x}}+C_1)(2\sigma_{x\hat{x}}+C_2)}{(\mu^2_x+\mu_{\hat{x}}^2+C_1)(\sigma_{x}^2+\sigma_{\hat{x}}^2+C_2)}
\end{equation}
where $\mu_x$ and $\mu_{\hat{x}}$ denote the means of images $\mathbf{x}$ and $\hat{\mathbf{x}}$, respectively;
$\sigma_x^2$ and $\sigma_{\hat{x}}^2$ are the variances;
$\sigma_{x\hat{x}}$ is the covariance;
and $C_1, C_2$ are constants. SSIM ranges from $-1$ to $1$, with larger values indicating higher similarity between the two images.

\begin{figure*}[t]
	\centerline{\includegraphics[width=7.2in,keepaspectratio]{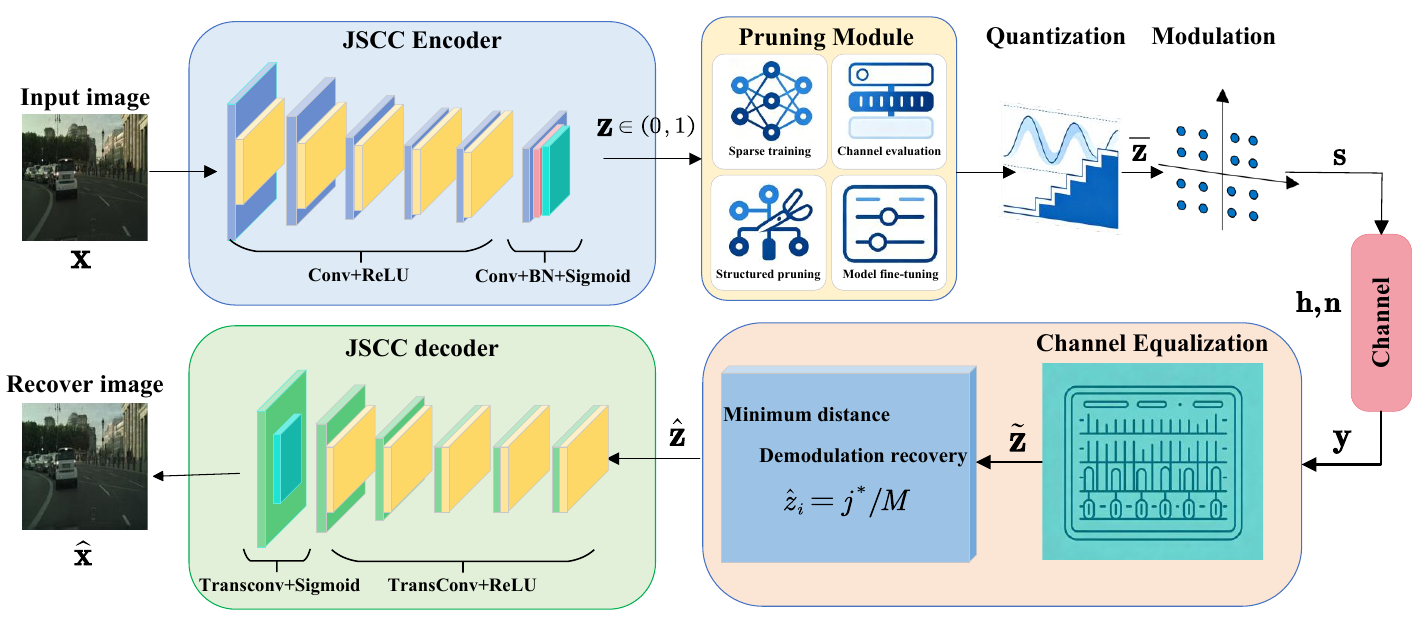}}
	\caption{Lightweight Low-SNR-Robust Semantic Communication Scheme.}
	\label{fig2}
\end{figure*}
\section{Proposed Scheme} \label{3}
To achieve lightweight, low-SNR robust semantic communication for autonomous driving, this paper proposes improvements from two aspects: model compression and transmission scheme design. The overall framework is shown in Fig. \ref{fig2}, which mainly includes: 1) Structured model pruning for in-vehicle platforms; 2) Quantization and modulation schemes adapted to JSCC encoded features.

\subsection{Lightweight Model Design Based on Structured Pruning}
The JSCC encoder and decoder employed in this paper are built upon deep convolutional neural networks.
The original model has a large number of parameters, making it difficult to directly deploy on resource-constrained vehicular computing platforms.
To address this, we introduce structured pruning to significantly reduce parameters and computational complexity while preserving model performance \cite{10204583}.

The core of structured pruning lies in evaluating and removing unimportant neurons or convolutional kernels in the network. Compared with unstructured pruning, structured pruning preserves the regular structure of the network, enabling practical inference acceleration and memory saving without requiring specialized hardware acceleration for the pruned model. This paper uses the scaling factors of BN layers to evaluate the importance of each channel. The transformation of the BN layer can be expressed as:
\begin{equation}
	\mathbf{z}_{\mathrm{out}}=\eta\frac{\mathbf{z}_{\mathrm{in}}-\mu_\mathcal{B}}{\sqrt{\sigma_{\mathcal{B}}^2+\epsilon}} + \beta,
\end{equation}
where $\mu_{\mathcal{B}}$ and $\sigma_{\mathcal{B}}^2$ are the mean and variance of the current batch input, respectively, and $\eta$ and $\beta$ are the learnable scaling factor and bias. The magnitude of the scaling factor $\eta$ directly reflects the activation strength of the corresponding channel \cite{Liu2017LearningEC}. Channels with small magnitudes contribute less to the network output and can thus be safely pruned.

To achieve channel sparsity, an L1 regularization term on $\eta$ is introduced into the training loss function. The overall loss function is defined as
\begin{equation}
	\mathcal{L}_{\mathrm{total}}=\underbrace{\mathbb{E}\left[\left\|\mathbf{x} - \hat{\mathbf{x}}\right\|_2^2\right]}_{\text{MSE reconstruction loss}} + \lambda\sum_{i=1}^{|\eta|}|\eta_i|,
\end{equation}
where the first term is the MSE between the original and reconstructed images, denoted as $\mathcal{L}_{\text{MSE}}$,
and the second term is the L1 regularization term with $\lambda$ as the regularization coefficient. During backpropagation, the regularization term drives unimportant $\eta_i$ toward zero.
Specifically, the gradient update rule for $\eta_i$ is given by:
\begin{equation}
	\eta_i^{\text{new}} = \eta_i^{\text{old}} - \alpha \left( \frac{\partial \mathcal{L}_{\text{MSE}}}{\partial \eta_i} + \lambda \cdot \text{sign}(\eta_i^{\text{old}}) \right),
\end{equation}
where $\alpha$ is the learning rate,
$\frac{\partial \mathcal{L}_{\text{MSE}}}{\partial \eta_i}$ is the partial derivative of the reconstruction loss with respect to $\eta_i$,
and $\text{sign}(\cdot)$ denotes the sign function. This update rule pushes small $\eta_i$ toward zero with a constant regularization gradient at each iteration, thus accelerating their sparsification.

\begin{algorithm}
	\SetAlgoLined
	\textbf{Initialization:} Pre-trained JSCC model $\mathcal{M}$, $\boldsymbol{\eta}$, training and validation set $\mathcal{D}_{\text{train}}$, $\mathcal{D}_{\text{val}}$, $\gamma$, $\lambda$, $E_{\text{sparse}}$, $E_{\text{pruning}}, $$E_{\text{fine-tune}}$\;
	
	\For{$1$ \KwTo $E_{\text{sparse}}$}{
		\For {each batch $(\mathbf{x}, \hat{\mathbf{x}}) \in \mathcal{D}_{\text{train}}$}
		{
			Compute total loss $\mathcal{L}_{\text{total}}$\;
			Backpropagate to obtain  $\frac{\partial \mathcal{L}_{\text{MSE}}}{\partial \boldsymbol{\theta}}$ and $\frac{\partial \mathcal{L}_{\text{MSE}}}{\partial \eta_i}$\;
			\For{each BN layer with $\eta_i$}{
				Update gradient by Equation (13)\;
			}
		}
		
		Update model parameters\;
	}
	\For{$1$ \KwTo $E_{\text{pruning}}$}{
		Removes least important channels according to current pruning ratio\;
		
		\For{$1$ \KwTo $E_{\text{fine-tune}}$}{
			Fine-tune $\mathcal{M}_t$ on $\mathcal{D}_{\text{train}}$ using only $\mathcal{L}_{\text{MSE}}$\;
			Evaluate on $\mathcal{D}_{\text{val}}$ and save the best model\;
		}
	}
	\caption{Structured Pruning Procedure}
	\label{alg:pruning}
\end{algorithm}

After sparse training, a global pruning rate $\gamma$ is set to remove the channels with the smallest $\gamma$ proportion of absolute $\eta$ values, along with the corresponding kernels in the adjacent convolutional layers. Such pruning slightly degrades model accuracy. Therefore, we fine-tune the pruned model for several epochs using the original training data with only the MSE loss $\mathcal{L}_{\text{MSE}}$,
enabling the model parameters to adapt to the new structure and recover reconstruction quality.
The complete structured pruning training pipeline is shown in Algorithm \ref{alg:pruning}.

\subsection{Low-SNR Robust Quantization Modulation Transmission}
Conventional semantic communication systems often adopt analog transmission, in which real or complex symbols directly output by the neural network encoder are fed into the channel. Although theoretically achieving optimal rate-distortion performance, this scheme is incompatible with existing mainstream digital communication systems and risks a cliff effect in low SNR regions. To address this, this paper designs a quantization and modulation scheme matched to JSCC encoded features, aiming to improve low-SNR robustness under digital transmission.

Since the output $\mathbf{z}$ of the JSCC encoder is constrained to $(0,1)$ by the Sigmoid function,
it is naturally compatible with a uniform quantizer.
We perform $M$-level uniform quantization on each $z_i$, as follows:
\begin{equation}
	\bar{z}_i = \frac{\lfloor z_i \times M \rfloor}{M}, i =1,2,...,k,
\end{equation}
where $\lfloor \cdot \rfloor$ denotes the floor operation,
and $M$ is the modulation order (e.g., 4, 16, 64, 256).
The quantized values $\bar{z}_i$ are evenly distributed over $[0, 1)$ with a total of $M$ distinct levels. Each $\bar{z}_i$ is then mapped to a unique $M$-QAM constellation point $s_i \in \mathcal{C}$,
where $\mathcal{C}$ denotes the constellation set.
The mapping can be expressed as:
\begin{equation}
	s_i=\mathcal{M}(\bar{z}_i)=\mathcal{C}_{\lfloor \bar{z}_i, M \rfloor},
\end{equation}
where $\mathcal{C}_j$ is the $j$-th symbol in the constellation set ($j = 0,1,\dots,M-1$).
The mapping is fully deterministic, and the average power of the constellation points remains constant due to the normalization of $\bar{z}_i$.

A key challenge of this scheme is that the quantization floor operation $\lfloor\cdot\rfloor$ is non-differentiable, preventing gradient backpropagation. To address this, we adopt a training-deployment separation strategy.
In the training phase, the quantization and modulation module is not introduced; the encoder output $\mathbf{z}$ is directly transmitted through the channel, and the received signal is
\begin{equation}
	\mathbf{y}_{\text{train}} = \mathbf{h} \odot \mathbf{z} + \mathbf{n}.
\end{equation}
After ideal channel equalization, the decoder input is
\begin{equation}
	\tilde{\mathbf{z}} = \mathbf{y}_{\text{train}} \oslash \mathbf{h} = \mathbf{z} + \mathbf{n} \oslash \mathbf{h}.
\end{equation}
The training loss is
$\mathcal{L}_{\text{MSE}} = \mathbb{E}[|\mathbf{x} - \mathcal{D}_\phi(\tilde{\mathbf{z}})|_2^2]$.
This process is fully differentiable, allowing gradients to backpropagate normally.

In the inference phase, the quantization function $\mathcal{Q}(\cdot)$ and modulation mapping $\mathcal{M}(\cdot)$ are applied.
The transmitted signal is $\mathbf{s} = \mathcal{M}(\mathcal{Q}(\mathbf{z}))$,
where $\mathcal{Q}$ quantizes each $z_i$ into $\bar{z}_i$,
and $\mathcal{M}$ maps $\bar{z}_i$ to the constellation point $s_i$. The received signal is
$\mathbf{y} = \mathbf{h} \odot \mathbf{s} + \mathbf{n}$.
The receiver performs minimum-distance demodulation to recover the quantized features:
\begin{equation}
	\hat{z}_i = \frac{1}{M} \cdot \arg\min_{j \in \{0,\dots,M-1\}} \|y_i - \mathcal{C}_j\|^2,
\end{equation}
That is,
$\hat{\mathbf{z}} = \mathcal{Q}^{-1}\big( \arg\min_j |\mathbf{y} - \mathcal{C}_j|^2 \big)$,
where $\mathcal{Q}^{-1}(j) = j/M$. Define the quantization error as $\boldsymbol{\epsilon}_q = \hat{\mathbf{z}} - \bar{\mathbf{z}}$, where $\bar{\mathbf{z}} = \mathcal{Q}(\mathbf{z})$;
and the channel demodulation error as $\boldsymbol{\epsilon}_c = \hat{\mathbf{z}} - \bar{\mathbf{z}}$, caused by noise and fading. The final input to the decoder is
$\hat{\mathbf{z}} = \bar{\mathbf{z}} + \boldsymbol{\epsilon}_c$.

Since the decoder has been adapted to noisy inputs $\tilde{\mathbf{z}} = \mathbf{z} + \mathbf{n}'$ during training (where $\mathbf{n}'$ denotes equivalent noise),
it inherently exhibits robustness against quantization and demodulation errors.
As long as the quantization is sufficiently fine (i.e., $M$ is large enough) such that $|\boldsymbol{\epsilon}_q| \ll |\mathbf{n}'|$,
the inference performance can approach that of training.
Experiments in the next section show that high-order QAM (e.g., 256QAM) suffices to satisfy this condition and outperforms conventional separate schemes at low SNR.
The complete pipeline of the above quantization-modulation and demodulation is illustrated in Algorithm \ref{alg:qam}.

\begin{algorithm}
	\SetAlgoLined
	
	\textbf{Initialization:} Original image $\mathbf{x}$, trained JSCC encoder $\mathcal{E}_{\boldsymbol{\theta}}$, decoder $\mathcal{D}_{\boldsymbol{\phi}}$, channel $\mathbf{h}$
	
	\textbf{Transmitter:}\;
	Extract semantic features: $\mathbf{z} = \mathcal{E}_{\boldsymbol{\theta}}(\mathbf{x}) \in (0,1)^k$\;
	Uniform quantization by Equation (14)\;
	Constellation mapping by Equation (15)\;
	Power normalization: $\mathbf{s} \leftarrow \sqrt{P} \cdot \mathbf{s} / \sqrt{\frac{1}{k}\sum_i |s_i|^2}$ to satisfy transmit power constraint\;
	Transmit $\mathbf{s}$ through channel: $\mathbf{y} = \mathbf{h} \odot \mathbf{s} + \mathbf{n}$\;
	
	\textbf{Receiver:}\;
	Channel equalization: $\tilde{\mathbf{z}} = \mathbf{y} \oslash \mathbf{h}$\;
	Minimum-distance demapping by Equation (18)\; 
	Compute $j^* = \arg\min_{j \in \{0,\dots,M-1\}} \|y_i - \mathcal{C}_j\|^2$\;
	Recover quantized features: $\hat{z}_i = j^* / M$\;
	Image reconstruction: $\hat{\mathbf{x}} = \mathcal{D}_{\boldsymbol{\phi}}(\hat{\mathbf{z}})$\;
	
	\caption{Quantization Modulation and Demodulation Inference Procedure}
	\label{alg:qam}
\end{algorithm}

\section{Simulation Experiments} \label{4}
This section verifies the performance of the proposed lightweight semantic communication system in autonomous driving scenarios through simulations.

\begin{table}
	\caption{The parameter settings of the model. Conv: Convolutional Layer. Trans ConV: Transposed convolutional layer.}\label{tab1}%
	\centering
	\begin{tabular}{@{}ccc@{}}
		\toprule
		Module&Layer Name&Kernel size, number of channels\\
		\midrule
		\multirow{6}{*}{Encoder}& Conv\tnote{a}  1 &(5, 5), 16\\	
		& Conv  2 & (5, 5), 32\\
		& Conv  3 & (5, 5), 64\\
		& Conv  4 & (5, 5), 128\\
		& Conv  5 & (5, 5), 256\\
		& Conv  6 with BN\tnote{b} & (5, 5), 512\\
		\midrule
		\multirow{6}{*}{Decoder}& Trans Conv\tnote{d}  1 &(5, 5), 256\\
		&Trans Conv  2 &(5, 5), 128\\
		&Trans Conv  3 &(5, 5), 64\\
		&Trans Conv  4 &(5, 5), 32\\
		&Trans Conv  5 &(5, 5), 16\\
		&Trans Conv  6 &(6, 6), 3\\
		\bottomrule
	\end{tabular}
\end{table}
\subsection{Simulation Parameter Settings}
Experiments are conducted on the Cityscapes dataset \cite{7780719} for model training and testing. This dataset contains finely annotated images of street scenes from 50 different cities, with a resolution of $1024 \times 2048$. In this paper, all images are resized to $512 \times 512$ to fit the network input.

\begin{figure*}[!t]
	\centering
	\subfloat[PSNR]{\includegraphics[width=3.0in]{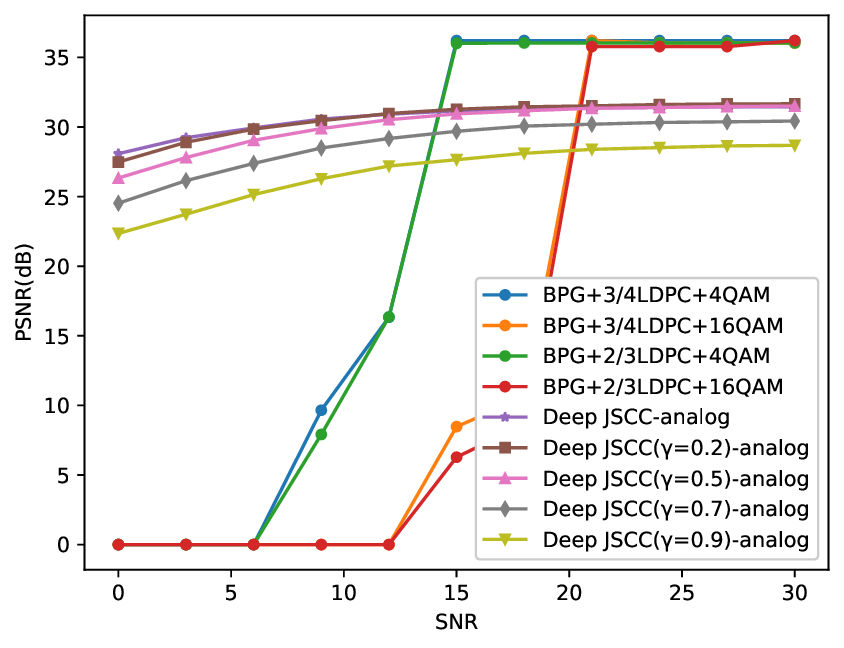}%
		\label{fig4a}}
	\hfil
	\subfloat[SSIM]{\includegraphics[width=3.0in]{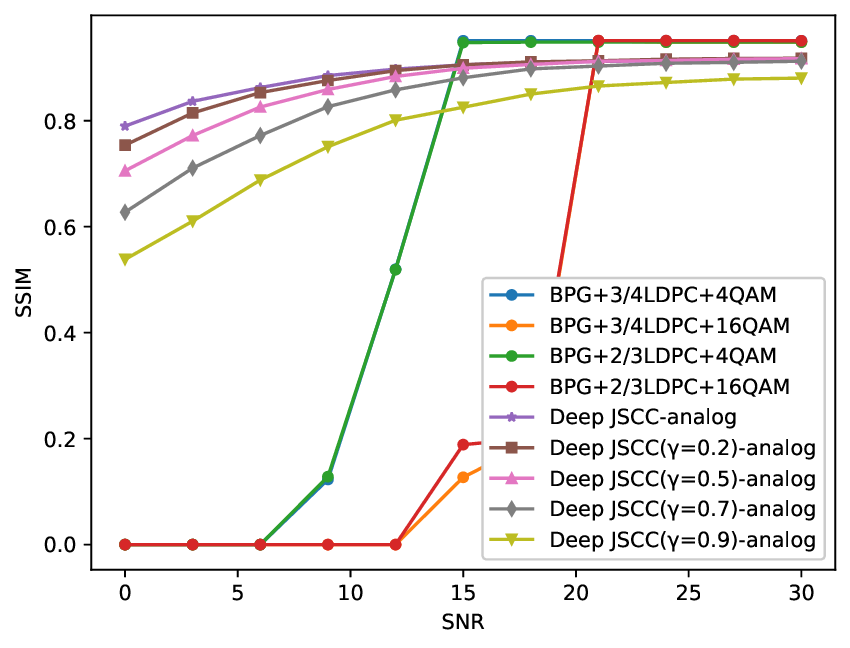}%
		\label{fig4b}}
	\caption{Comparison of the performance of the proposed scheme under different pruning ratios with the traditional BPG-LDPC transmission scheme, where (a) and (b) are the results obtained under slow Rayleigh fading channels.}
	\label{fig4}
\end{figure*}
\begin{figure*}[!h]
	\centering
	\subfloat[PSNR]{\includegraphics[width=3.0in]{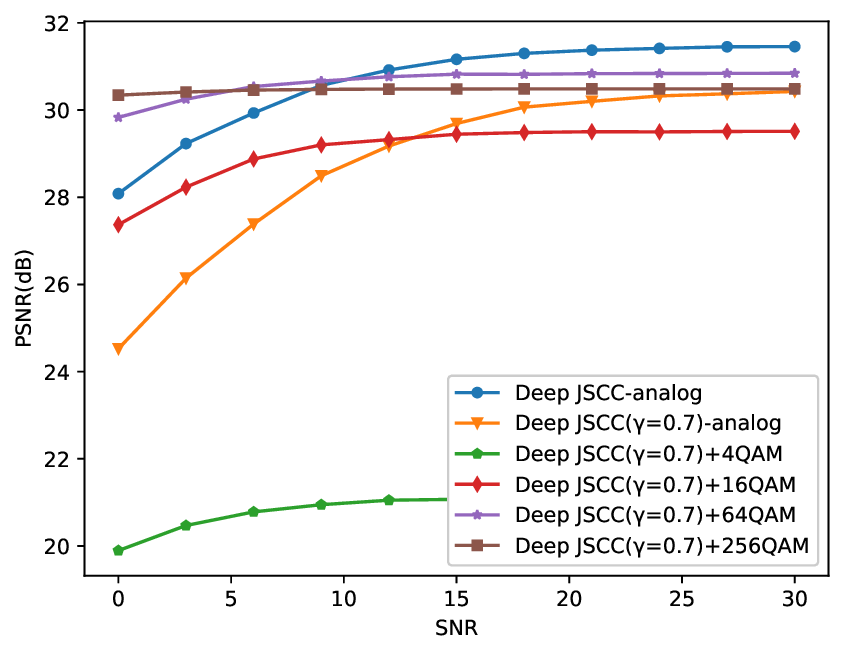}%
		\label{fig5a}}
	\hfil
	\subfloat[SSIM]{\includegraphics[width=3.0in]{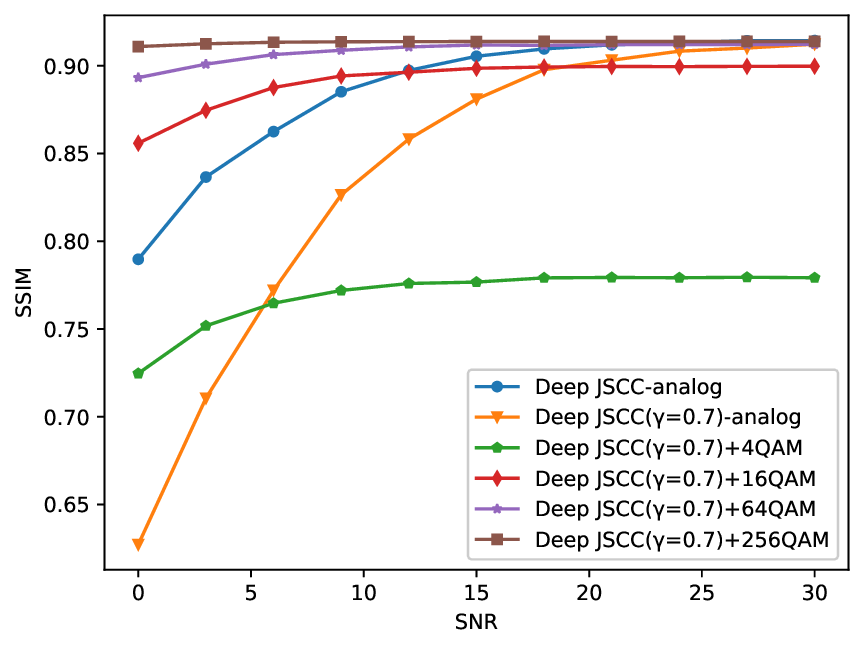}%
		\label{fig5b}}
	\caption{Comparison of the performance of the proposed deep JSCC when pruning ratio is set to $\gamma=0.7$ with the traditional BPG-LDPC using QAM transmission, where (a) and (b) are the results obtained under slow Rayleigh fading channels.} 
	\label{fig5}
\end{figure*}

The JSCC encoder and decoder are trained jointly in an end-to-end manner.
The number of training epochs is $N_{\text{sparse}} = 10$ and $N_{\text{pruning}} = 4$.
The batch size is $B = 32$, the initial learning rate is $\alpha = 1\times10^{-4}$,
and the optimizer is Adam. The signal-to-noise ratio during training is fixed at $25$ dB, and the channel model is a slow Rayleigh fading channel.
The bandwidth compression ratio is set to $k/n = 2/3$, meaning that each pixel is transmitted using an average of $2/3$ channel symbols  \cite{8723589}.
The pruning rate $\gamma$ is set to 0, 0.2, 0.5, 0.7, and 0.9, respectively. In sparse training, the L1 regularization coefficient is $\lambda = 1\times10^{-5}$,
and the fine-tuning epochs after pruning are $N_{\text{fine-tune}} = 5$.
All pruning and training processes are performed on the server side,
while only the pruned lightweight model is loaded on the vehicle side for inference. PSNR and SSIM are used to evaluate the quality of reconstructed images.
All results are obtained by averaging over the test set.
The remaining parameter settings are shown in Table \ref{tab1}.

To verify the superiority of the proposed scheme, we compare it with the conventional separate scheme using QAM modulation + LDPC channel coding + BPG image coding.
BPG is a high-efficiency still image coding standard. The code rate of the LDPC code is $3/4$;
modulation schemes include 4QAM, 64QAM, and 256QAM.
In the conventional scheme, source coding and channel coding are designed independently, following the Shannon separation theorem.

\begin{figure*}[h]
	\centering
	\begin{minipage}{0.20\linewidth}
		\centerline{Original}
		\vspace{3pt}
		\centerline{\includegraphics[scale=0.15]{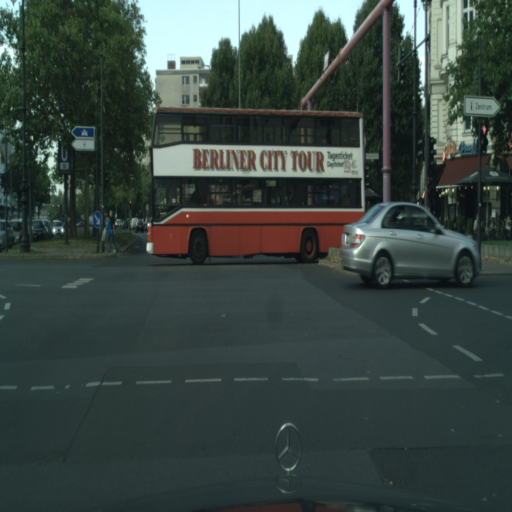}}
		\centerline{PSNR~/~SSIM}
	\end{minipage}
	\hspace{-0.5cm}
	\begin{minipage}{0.20\linewidth}
		\centerline{$\gamma=0.7$-analog}
		\vspace{3pt}
		\centerline{\includegraphics[scale=0.15]{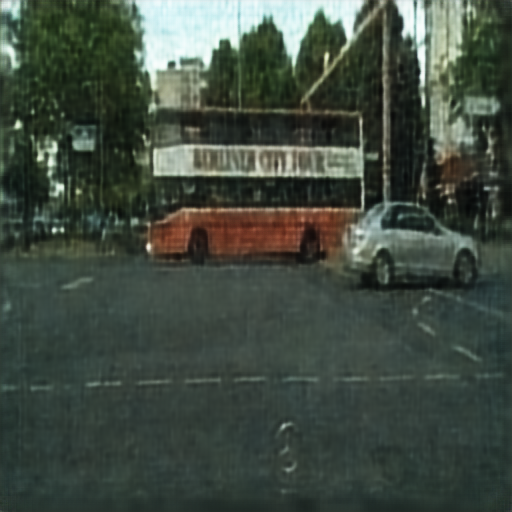}}
		\centerline{26.64 dB~/~0.84}
	\end{minipage}
	\hspace{-0.5cm}
	\begin{minipage}{0.20\linewidth}
		\centerline{$\gamma=0.7$+256QAM}
		\vspace{3pt}
		\centerline{\includegraphics[scale=0.15]{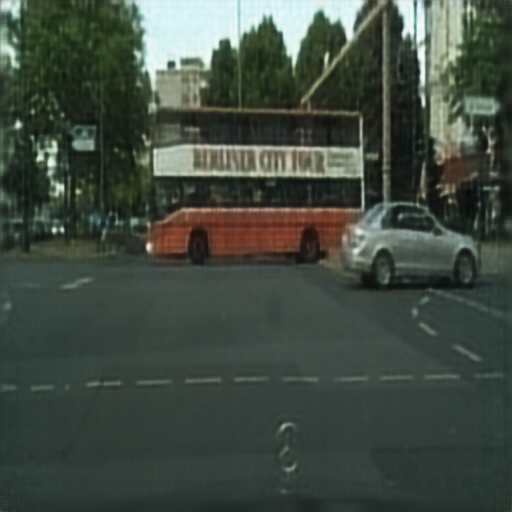}}
		\centerline{27.76 dB~/~0.90}
	\end{minipage}
	\hspace{-0.5cm}
	\begin{minipage}{0.20\linewidth}
		\centerline{BPG+LDPC+QAM}
		\vspace{3pt}
		\centerline{\phantom{\includegraphics[scale=0.15]{Figures/Fig6_first_column.png}}}
		\centerline{N~/~A}
	\end{minipage}
	
	\begin{minipage}{0.20\linewidth}
		\vspace{3pt}
		\centerline{\includegraphics[scale=0.15]{Figures/Fig6_first_column.png}}
		\centerline{PSNR~/~SSIM}
	\end{minipage}
	\hspace{-0.5cm}
	\begin{minipage}{0.20\linewidth}
		\vspace{3pt}
		\centerline{\includegraphics[scale=0.15]{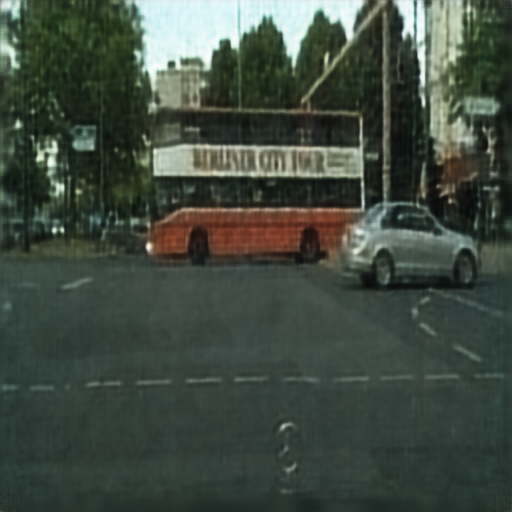}}
		\centerline{27.39 dB~/~0.88}
	\end{minipage}
	\hspace{-0.5cm}
	\begin{minipage}{0.20\linewidth}
		\vspace{3pt}
		\centerline{\includegraphics[scale=0.15]{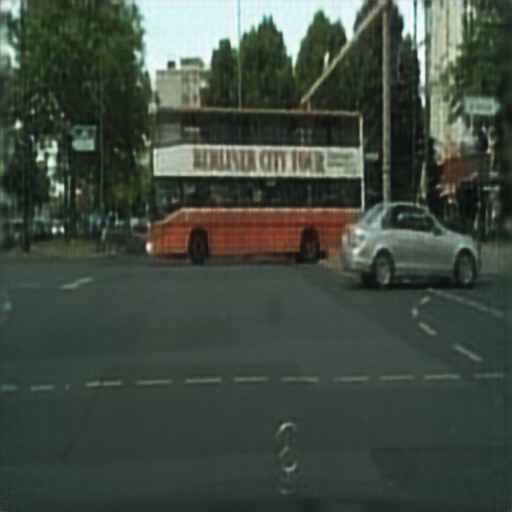}}
		\centerline{27.80 dB~/~0.90}
	\end{minipage}
	\hspace{-0.5cm}
	\begin{minipage}{0.20\linewidth}
		\vspace{3pt}
		\centerline{\includegraphics[scale=0.15]{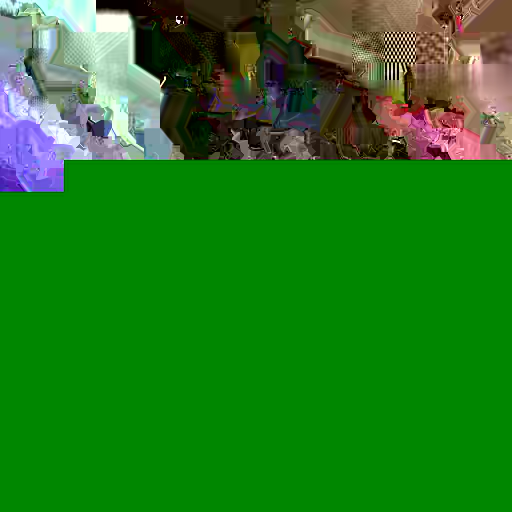}}
		\centerline{9.50 dB~/~0.13}
	\end{minipage}
	
	\begin{minipage}{0.20\linewidth}
		\vspace{3pt}
		\centerline{\includegraphics[scale=0.15]{Figures/Fig6_first_column.png}}
		\centerline{PSNR~/~SSIM}
	\end{minipage}
	\hspace{-0.5cm}
	\begin{minipage}{0.20\linewidth}
		\vspace{3pt}
		\centerline{\includegraphics[scale=0.15]{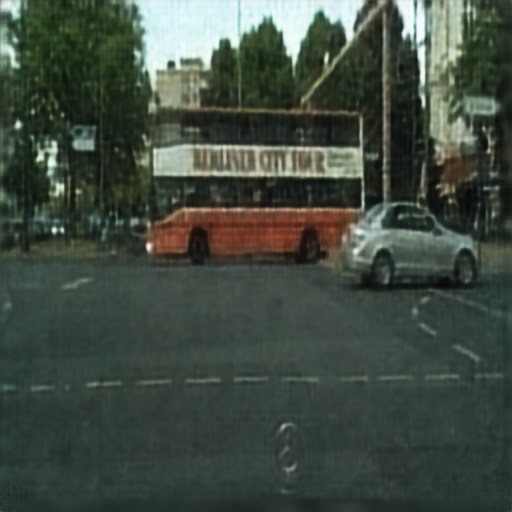}}
		\centerline{27.20 dB~/~0.87}
	\end{minipage}
	\hspace{-0.5cm}
	\begin{minipage}{0.20\linewidth}
		\vspace{3pt}
		\centerline{\includegraphics[scale=0.15]{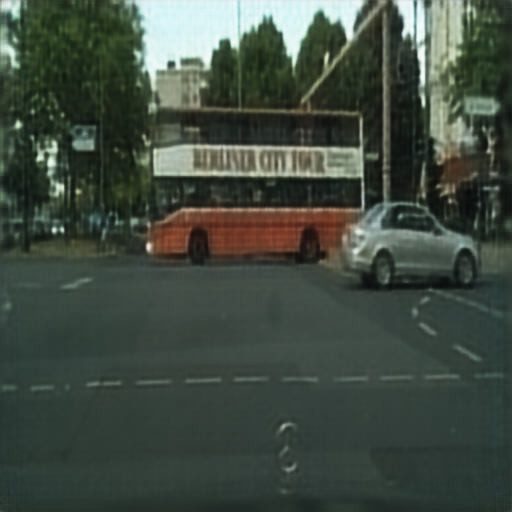}}
		\centerline{27.80 dB~/~0.90}
	\end{minipage}
	\hspace{-0.5cm}
	\begin{minipage}{0.20\linewidth}
		\vspace{3pt}
		\centerline{\includegraphics[scale=0.15]{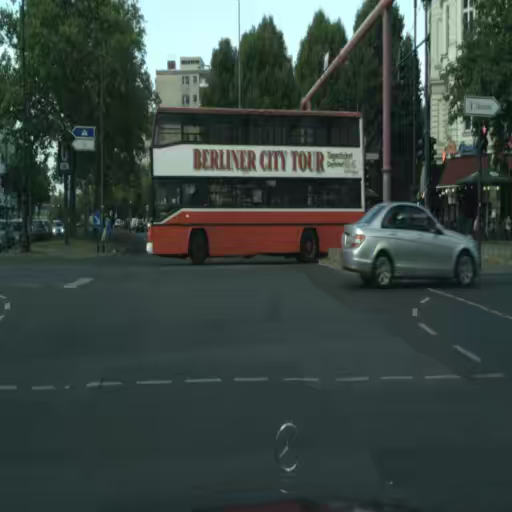}}
		\centerline{35.27 dB~/~0.95}
	\end{minipage}
	
	\begin{minipage}{0.20\linewidth}
		\vspace{3pt}
		\centerline{\includegraphics[scale=0.15]{Figures/Fig6_first_column.png}}
		\centerline{PSNR~/~SSIM}
	\end{minipage}
	\hspace{-0.5cm}
	\begin{minipage}{0.20\linewidth}
		\vspace{3pt}
		\centerline{\includegraphics[scale=0.15]{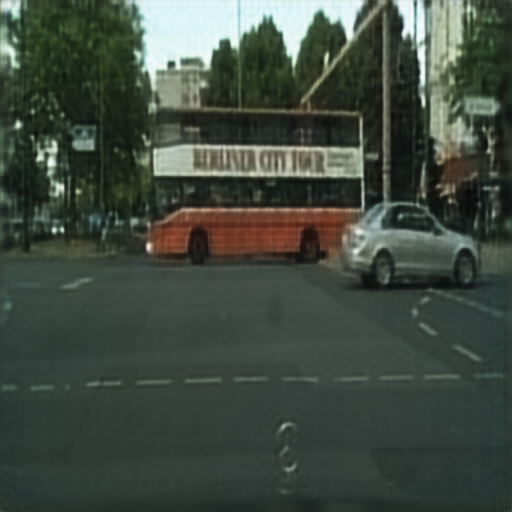}}
		\centerline{27.91 dB~/~0.91}
	\end{minipage}
	\hspace{-0.5cm}
	\begin{minipage}{0.20\linewidth}
		\vspace{3pt}
		\centerline{\includegraphics[scale=0.15]{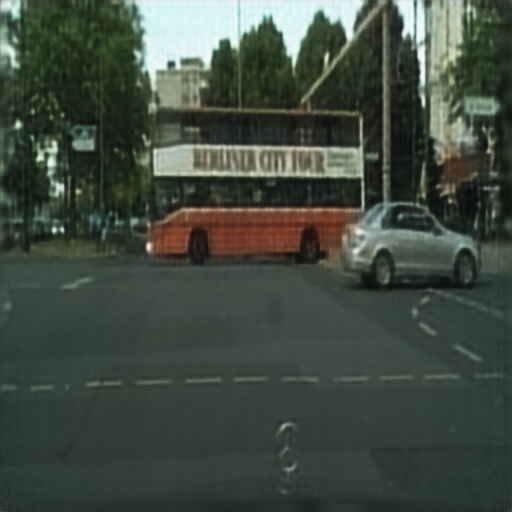}}
		\centerline{27.80 dB~/~0.90}
	\end{minipage}
	\hspace{-0.5cm}
	\begin{minipage}{0.20\linewidth}
		\vspace{3pt}
		\centerline{\includegraphics[scale=0.15]{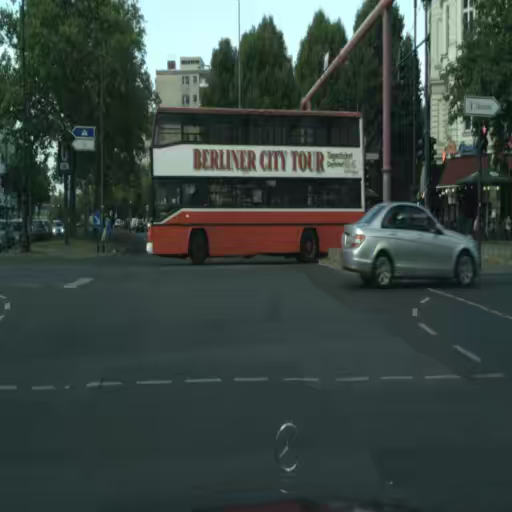}}
		\centerline{35.27 dB~/~0.95}
	\end{minipage}
	\caption{Examples of reconstructed images generated from the proposed scheme and the baseline digital scheme, with the baseline scheme using BPG-LDPC with QAM for transmission over slow Rayleigh channels. The SNR values of $1$ dB, $9$ dB, $15$ dB, and $22$ dB correspond to the rows arranged in ascending order from top to bottom.}
	\label{fig6}
\end{figure*}

\subsection{Performance Comparison with Separate Coding Schemes}
Fig. \ref{fig4} presents the performance comparison between the proposed deep JSCC model and its pruned versions against the BPG-LDPC scheme over a slow Rayleigh fading channel. Deep JSCC-analog denotes the analog transmission scheme that feeds signals directly into the channel. It can be observed from Fig.~\ref{fig4}(a) that in the high SNR region, the PSNR of the BPG-LDPC scheme is higher than that of the deep JSCC scheme, which is a typical advantage of separate coding under favorable channel conditions. When the SNR is below 14.5 dB, the trend reverses: the PSNR of the deep JSCC scheme is significantly superior to that of the BPG-LDPC scheme. Due to LDPC decoding failures, the PSNR of the BPG scheme drops sharply to nearly 0 dB, exhibiting the classic cliff effect.
When $\gamma = 0.2$, its PSNR curve almost coincides with that of the unpruned model.
Even when $\gamma = 0.5$ (half of the parameters are pruned), the PSNR loss is negligible across the entire SNR range.
This indicates that there exist numerous redundant neurons in the deep JSCC model, and structured pruning can safely remove them with almost no performance degradation.

The SSIM results in Fig.~\ref{fig4}(b) exhibit the same trend.
When the SNR is above 14.8 dB, the BPG scheme is superior;
in the low-SNR region, however, the SSIM values of the deep JSCC models (including pruned versions) are significantly higher than those of the BPG scheme.
Notably, the pruned model with $\gamma = 0.5$ even achieves slightly higher SSIM than the unpruned model at some SNR points,
which may be attributed to the regularization effect introduced by pruning that improves the generalization ability of the model.

\subsection{Robustness Analysis Under Different Modulation Orders}
Fig. \ref{fig5} shows the performance of the pruned model ($\gamma = 0.7$) under different modulation orders. As a benchmark, the results of analog transmission are also provided.
It can be seen from the PSNR curves in Fig.~\ref{fig5}(a) that Deep JSCC-analog achieves the best performance across the entire SNR range, since it avoids quantization errors. When larger QAM is adopted, the quantization error is minimized, and the PSNR curve is very close to that of analog transmission, with a gap of less than 0.5 dB above SNR = 15 dB. As the modulation order decreases, the quantization error increases and the PSNR decreases accordingly. However, even in the worst case of 4QAM, the PSNR remains above 20.4 dB at SNR = 15 dB.

The SSIM results in Fig.~\ref{fig5}(b) further confirm the above trend.
Importantly, under extremely low SNR conditions, the SSIM values of all modulation schemes remain above 0.7,
whereas the SSIM of the BPG scheme in Fig.~\ref{fig4}(b) has already dropped to 0 at this SNR. This fully demonstrates the robustness of the proposed quantized modulation scheme in low SNR environments:
benefiting from the noise robustness learned by the JSCC decoder during the analog training phase, the decoder can still recover recognizable image content from disturbed quantized features even if partial errors occur during demodulation.

Fig. \ref{fig6} shows examples of reconstructed images for each scheme under different SNR conditions (1 dB, 9 dB, 15 dB, 22 dB). It can be observed that the BPG scheme completely fails at SNR = 1 dB and 9 dB, and the reconstructed images suffer from severe block artifacts and color distortion, which is a direct manifestation of the cliff effect. 
Deep JSCC ($\gamma=0.7$-analog) maintains the basic contours of images at low SNR, but exhibits salt-and-pepper-like graininess with considerable loss of detailed textures. 
Deep JSCC ($\gamma=0.7$ + 256QAM) can still produce visually acceptable images at SNR = 1 dB, with a PSNR of 27.76 dB and an SSIM of 0.90.
Compared with analog transmission, the discrete constellation mapping introduced by quantization modulation exerts a certain denoising effect, resulting in cleaner reconstructed images with sharper contours. This phenomenon indicates that hard decisions in digital transmission can suppress the minor noise disturbances accumulated in analog transmission under certain conditions.

\begin{figure*}[t]
	\centering
	\subfloat[PSNR]{\includegraphics[width=3.0in]{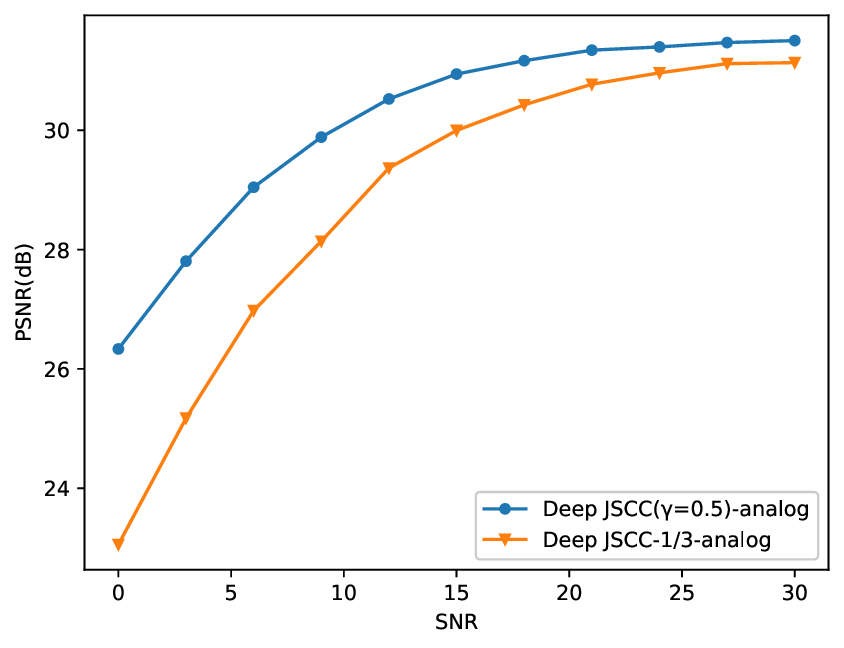}%
		\label{fig7a}}
	\hfil
	\subfloat[SSIM]{\includegraphics[width=3.0in]{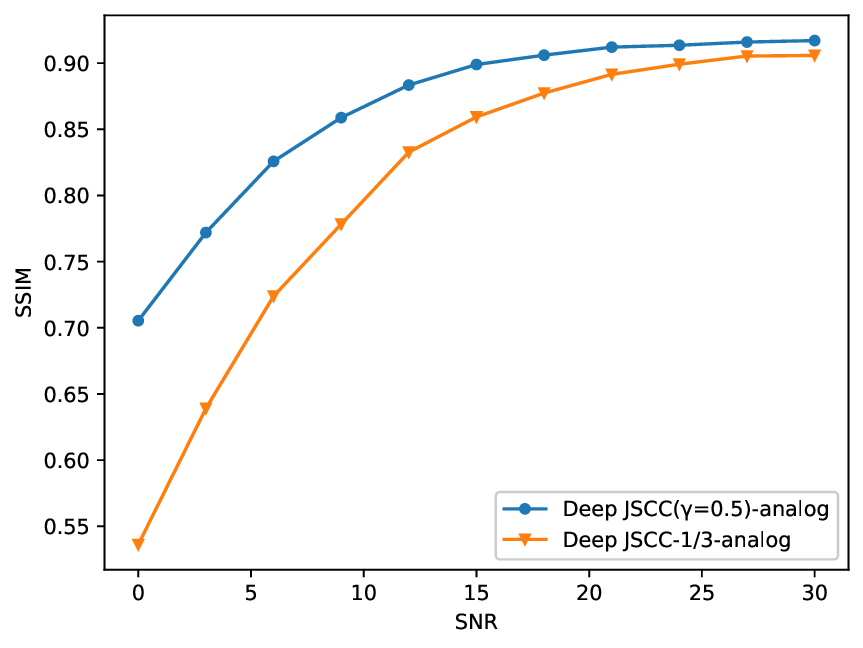}%
		\label{fig7b}}
	\caption{Comparison of performance between two different compressed models, where the results of (a) and (b) were obtained under the training condition of SNR=$25$ dB.}
	\label{fig7}
\end{figure*}

\begin{table*}[t]
	\caption{Comparison of the parameter and computational requirements for the proposed with different pruning ratios $\gamma$ at SNR = 25 {\upshape dB}.}\label{tab2}
	\centering
	\begin{tabular*}{\textwidth}{@{\extracolsep\fill}lllllll}
		\toprule
		Scheme& Pruning ratio& Parameters (M) & Size (M) &  MACs (G) & PSNR & SSIM\\
		\midrule
		\multirow{5}{*}{Proposed}&$\gamma=0$ & 6.25 & 23.8 & 20.24& 31.42 & 0.91 \\	
		&$\gamma=0.2$ & 5.36& 20.4 & 19.35& 31.63& 0.92\\
		&$\gamma=0.5$ & 4.02 & 15.3 & 18.03 &30.99 & 0.91\\	
		&$\gamma=0.7$ & 3.12 & 11.9 & 17.14 & 30.36 & 0.91\\	
		&$\gamma=0.9$ & 2.24 & 8.54 & 16.26 & 28.57 & 0.87\\
		\multirow{1}{*}{BPG+LDPC+64QAM}&/ & / & / & / & 35.55 &0.95\\
		\bottomrule
	\end{tabular*}
\end{table*}

\subsection{Validation of the Effectiveness of the Pruning Strategy}
Fig. \ref{fig7} compares the performance of the pruned model ($\gamma=0.5$, original bandwidth compression ratio $k/n=2/3$) with a model obtained by directly reducing the bandwidth compression ratio ($k/n=1/3$, with a similar number of parameters to the pruned model). The results show that under the same number of parameters, both PSNR and SSIM of the pruned model are significantly better than those of the bandwidth-compressed model.
This is because the pruning algorithm can intelligently retain important channels and remove redundant ones, whereas simply reducing the compression ratio uniformly weakens the representation ability of all channels, leading to more severe performance degradation.
This demonstrates the superiority of structured pruning in model lightweighting.

\subsection{Model Complexity Analysis}
Table \ref{tab2} lists the number of parameters, multiply–accumulate operations (MACs), and average PSNR/SSIM at SNR = 25 dB for the deep JSCC model under different pruning ratios. When $\gamma=0$, it corresponds to the unpruned deep JSCC model, with 7.19 M parameters and 28.94 G MACs.
At $\gamma=0.5$, both the number of parameters and MACs are nearly halved, and as shown in Fig.~\ref{fig4}, Fig.~\ref{fig5}, and Fig.~\ref{fig6}, the model still maintains high PSNR and SSIM performance. In addition, Table \ref{tab2} presents the average performance of each model on the Cityscapes dataset at SNR = 25 dB. Except for the scheme with a pruning ratio of 0.9, all other pruned schemes maintain a PSNR above 30 dB and an SSIM above 0.9.

\section{Conclusion}  \label{5}
To address key issues in autonomous driving V2V collaborative perception scenarios, such as limited on-board resources, time-varying fading wireless channels, and insufficient low-SNR robustness, this paper proposed a lightweight semantic communication system with strong low-SNR robustness. Based on deep JSCC as the core, the system was designed from two aspects: model lightweighting and transmission robustness. It effectively overcame the cliff effect of traditional separate communication schemes, as well as the shortcomings of existing semantic communication models such as high complexity and poor digital compatibility. Extensive simulation experiments based on the Cityscapes dataset verified that the reconstruction quality of the proposed system in the low SNR region was significantly superior to the traditional BPG-LDPC separate scheme; its high-order modulation performance approached that of analog transmission, and the pruned model outperformed the direct bandwidth compression model under the same number of parameters.

\bibliographystyle{ieeetr} 
\bibliography{MyRefs} 

\end{document}